\documentclass[
superscriptaddress,
 amsmath,amssymb,
 aps,
pra,
twocolumn,%Two column formatting.
]{revtex4-1}
\usepackage{graphicx}% Include figure files
\usepackage{dcolumn}% Align table columns on decimal point
\usepackage{bm}% bold math
\usepackage{gensymb}
\usepackage{bm,physics}
\usepackage{stackengine}

\usepackage[usenames, dvipsnames]{color}
\usepackage{float}   % for fig to beinside text
\usepackage{hyperref} % PDF links
\hypersetup{colorlinks=true,citecolor=blue,linkcolor=blue,urlcolor=blue}

\begin{document}

%\preprint{APS/123-QED}

\title{Photon-photon Correlations from a Pair of Strongly Coupled Two-Level Emitters}% Force line breaks with \\

\author{Elnaz Darsheshdar}
\email{darsheshdare@gmail.com}
\affiliation{%
Departamento de F\'{i}sica, Universidade Federal de S\~{a}o Carlos,
P.O. Box 676, 13565-905, S\~{a}o Carlos, S\~{a}o Paulo, Brazil
}
%Departamento de F´ısica, Universidade Federal de S˜ao Carlos,
%P.O. Box 676, 13565-905, S˜ao Carlos, S˜ao Paulo, Brazil 
\author{Mathilde Hugbart}
\email{mathilde.hugbart@inphyni.cnrs.fr}
\affiliation{%
Universit\'e C\^ote d'Azur, CNRS, INPHYNI, France%
}

\author{Romain Bachelard}
\email{bachelard.romain@gmail.com}
\affiliation{%
Departamento de F\'{i}sica, Universidade Federal de S\~{a}o Carlos,
P.O. Box 676, 13565-905, S\~{a}o Carlos, S\~{a}o Paulo, Brazil
}

\author{Celso Jorge Villas-Boas}
\email{celsovb@df.ufscar.br}
\affiliation{%
Departamento de F\'{i}sica, Universidade Federal de S\~{a}o Carlos,
P.O. Box 676, 13565-905, S\~{a}o Carlos, S\~{a}o Paulo, Brazil
}

%\affiliation{
%\affiliation{
%\affiliation{%


\date{\today}% It is always \today, today,

\begin{abstract}
We investigate two-color photon correlations in the light emitted by strongly coupled two-level emitters. Spectral filtering allows us to manipulate the collected light statistics and we show that the resonances induced by dipole-dipole interactions give rise to specific correlations, where the time-symmetry of the correlations is broken. Based on the collective dressed states, our study encompasses both the case of real processes, where the photons are associated with specific resonances and classical correlations between each other, and virtual processes, where pairs of photons are emitted with non-classical correlations.
\end{abstract}

\maketitle

%\tableofcontents

\section{Introduction}

Two-level emitters essentially behave as classical oscillators when they are weakly driven, and the light elastically scattered by such systems presents a range of optical phenomena that can be understood using the tools of linear optics: dispersion, Rayleigh scattering, but also cooperative phenomena such as superradiance~\cite{Agarwal1974}. Differently, for a single emitter, a strong drive results in a strong inelastic component in the scattering, with the emergence of a ``Mollow triplet'' composed of a carrier centered at the laser frequency, and two symmetric sidebands shifted away by the Rabi frequency of the driving field~\cite{Mollow}. Of particular interest are the strong correlations between the photons emitted from the two sidebands: originally measured in atomic systems~\cite{Aspect} and extensively studied theoretically~\cite{Cohen,Dalibard,Schrama}, the field has known a resurgence of attention in the context of quantum dots, with a recent measurement of such photon (anti)correlations between sidebands~\cite{Ulhaq}, thus demonstrating the potential of artificial atoms as sources of heralded photons. 

In this context, the coupling of emitters gives access to new control parameters, as interaction-induced resonances rise, but also interferences phenomena~\cite{Wolf2020}. Indeed, the coupling of the emitters through common radiation modes result in the commonly called dipole-dipole interactions, which manifests in both the exchange of excitations and cooperative decay processes~\cite{Lehmberg1970a,Lehmberg1970b}. As a consequence, the fluorescence spectrum of strongly driven atoms presents new sidebands~\cite{Senitzky,Agarwall,Aryeh} which, for a weak interaction, appear at twice the Rabi frequency from the carrier. Such effects are expected to show up, for instance, in many-atom extended cloud, with the resonant optical thickness playing the role of cooperativity parameter~\cite{Pucci}.

Nevertheless, the diversity of photon-photon correlations emitted from strongly-interacting systems has barely been scratched. The recent development of the so-called sensor method~\cite{delValle}, where the photons emitted in a given mode are monitored by introducing an artificial two-level system resonant with its frequency (analogously to a Fabry-Perot cavity), has allowed to explore extensively multi-photon correlations for single emitters~\cite{LopezCarreno}. In particular, the potential of virtual transitions, where photons are emitted in bundles, as a source of quantum correlations has been pointed out. As for interacting emitters, the quantum correlations which emerge for two weakly interacting emitters have been investigated in the specific configuration of a pump driving a single emitter, although the fluorescence spectrum is not substantially affected by the interaction in this configuration~\cite{Peng_2019}.  

In this work, we investigate two-color correlations in the light emitted by two strongly-driven strongly-interacting emitters. The correlations are interpreted introducing the collective dressed states 
picture, which allows us to describe both bunching and anti-bunching based on the transitions between them. At odds from weakly-interacting emitters, the strong interaction lifts the degeneracy of the energy differences between the different states 
leading to a temporal breaking of symmetry for the correlations: photons of different frequencies may not be emitted in any order. Finally we show that most of the virtual processes, which involve pairs of photons, yield non-classical correlations when the sum of their energies fits any of the interaction-induced sidebands.

\section{Modeling and detection scheme}

In our work, we consider two identical two-level systems. Experimentally, this can be either two atoms, two molecules or two quantum dots. For this last example, although quantum dots are promising single-emitter platforms, it remains very challenging to produce very similar dots, i.e., with very similar transition frequencies and linewidths. On the other hand, laser cooling has allowed to bring the interactions between cold atoms under a very high degree of control. In this paper, the two-level system is considered to be a motionless atom.

The system under study is thus composed of two identical driven two-level systems (TLS) at positions ${{\textbf{r}}_{i}}$, with transition frequency $\omega_a$ and linewidth $\Gamma$. Each atom is described by the spin-half angular momentum algebra, with $\sigma _{i}^{-}$ ($\sigma _{i}^{+}$) the lowering (rising) operator of the $i$th atom ($i=1,2$). In the Born, Markov and rotating-wave approximations, the master equation which describes the dynamics of its density matrix $\rho$, in the laser reference frame, is given by \cite{Agarwal} (we set $\hbar =1$ along the paper):
\begin{equation}
\frac{\partial \rho }{\partial t}=i\left[ \rho ,H \right]+{\mathcal{L}}\rho,
\label{mast}
\end{equation}
where the coherent and incoherent parts are encoded in the following at resonance Hamiltonian and Lindblad super-operator, respectively: 
\begin{eqnarray}
H&=&\frac{1}{2}\sum\limits_{i}[\Omega^*(\mathbf{r}_j)\sigma _{i}^{-}+\Omega(\mathbf{r}_j)\sigma _{i}^{+}]+\Gamma\sum\limits_{i,j\ne i}{{\delta }_{ij}}\sigma _{i}^{+}\sigma _{j}^{-},
\label{Ham}
\\ \mathcal{L}\rho &=&\frac{\Gamma }{2}\sum\limits_{i}(2\sigma _{i}^{-}\rho \sigma _{i}^{+}-\sigma _{i}^{+}\sigma _{i}^{-}\rho -\rho \sigma _{i}^{+}\sigma _{i}^{-}) \nonumber 
\\
&& +\frac{\Gamma }{2}\sum\limits_{i,j\ne i}{{\gamma }_{ij}}(2\sigma _{j}^{-}\rho \sigma _{i}^{+}-\sigma _{i}^{+}\sigma _{j}^{-}\rho -\rho \sigma _{i}^{+}\sigma _{j}^{-}). 
\label{Lin}
\end{eqnarray}
The atoms are here resonantly driven by a monochromatic plane--wave $\Omega(\mathbf{r})=\Omega e^{i\mathbf{k}_L.\mathbf{r}}$, where $\Omega$ stands for the Rabi frequency and $\mathbf{k}_L$ the light wavevector. The dipole-dipole interactions give rise to both a coherent and incoherent coupling
\begin{eqnarray}
\label{gamma}
&&{{\delta }_{ij}}=-\frac{3}{4}\left( 1-\cos^{2} {{ \theta_{ij}}} \right)\frac{\cos{(k{r}_{ij})}}{{k{r}_{ij}}}
\\ &&+\frac{3}{4}\left( 1-3\cos^{2} {{\theta_{ij}}} \right)\left[ \frac{\sin {(k{r}_{ij})}}{(kr_{ij})^{2}}+\frac{\cos {(k{r}_{ij})}}{(kr_{ij})^{3}} \right],\nonumber
\\ &&{{\gamma }_{ij}}=\frac{3}{2}\left( 1-\cos^{2} {{ \theta_{ij}}} \right)\frac{\sin{(k{r}_{ij})}}{{k{r}_{ij}}}\nonumber
\\ &&+\frac{3}{2}\left( 1-3\cos^{2} {{\theta_{ij}}} \right)\left[ \frac{\cos{(k{r}_{ij})}}{(kr_{ij})^{2}}-\frac{\sin{(k{r}_{ij})}}{(kr_{ij})^{3}} \right],\nonumber
\end{eqnarray}
with $\lambda=2\pi/k$ the wavelength transition ($k\approx k_L$), $r_{ij}$ the distance between the atoms, and $\theta_{ij}$ the angle between their dipole moments and the vector joining them, $\mathbf{r}_{ij}=\mathbf{r}_{j}-\mathbf{r}_{i}$.

Solving the master equation provides the scattered electric field, which is given, in the far field limit and in direction $ \hat{\mathbf{n}}$, by 
\begin{equation}
    {{E}^{\dagger }}\left(t\right)=\sum\limits_{j=1}^{2}{\sigma _{j}^{-}\left( t \right){{e}^{-ik\mathbf{\hat{n}}.{{\mathbf{r}}_{j}}}}}.\label{eq:E}
\end{equation}
The dependence on $ \hat{\mathbf{n}}$ is hereafter kept implicit. 
Its temporal coherence is captured by the first-order and second-order two-time correlation functions:
\begin{eqnarray}
\label{eq2} 
{g}^{(1)}\left(\tau  \right)&=&\lim_{t\to\infty}\frac{\left\langle E\left( t \right){{E}^{\dagger }}\left( t+\tau  \right) \right\rangle }{\left\langle E\left(t \right){{E}^{\dagger }}\left(t \right) \right\rangle }, 
\\ {g}^{(2)}\left(\tau  \right)&=&\lim_{t\to\infty}\frac{\left\langle E\left( t \right) E\left( t+\tau \right){{E}^{\dagger }}\left( t+\tau  \right) {{E}^{\dagger }}\left( t  \right) \right\rangle }{\left\langle E\left(t \right){{E}^{\dagger }}\left(t \right) \right\rangle^2 },\label{eq:defg2}
\end{eqnarray}
here computed in the steady state. In particular, the fluorescence spectrum, sometimes referred to one-photon spectrum (1PS), is obtained from the Fourier transform of the first-order correlation function
\begin{equation} 
\label{eq:1PS}
S\left( \omega  \right)=\underset{T\to \infty }{\mathop{\lim }}\,\int_{-T}^{T}{ {{g}^{(1)}}\left(\tau  \right){{e}^{-i\omega \tau }}d\tau}.
\end{equation}
The 1PS gives the spectral energy distribution of the light scattered elastically and inelastically, whereas the second-order correlation function $g^{(2)}$ contains details on the correlations between the emitted photons, with the antibunching in the trains of photons emitted by a single emitter as a hallmark of the non-classicality of this emission~\cite{Kimble1977}.

The problem of time-resolved observables is however more challenging. Indeed, as one introduces the field operator in the reciprocal space $\tilde{E}(\omega)=\int_{t=-\infty}^\infty e^{-i\omega t}E(t)dt $, the problem of studying two-color photon-photon correlations brings in the calculation of a four-time correlator, $\left\langle E(t_1) E(t_2) {E}^{\dagger }(t_3){E}^{\dagger}(t_4) \right\rangle $. Then, the use of the quantum regression theorem, commonly used for two-time observables, may become a daunting task~\cite{Gisin1993,Brun1996,Breuer1997}. This is a strong restriction to the study of photon-photon correlations, which has long limited rigorous results to single-emitter physics~\cite{Bel2009}.

An elegant solution was found in the “sensor method” that allows one to investigate theoretically frequency-resolved correlations in greater details~\cite{delValle}. It relies on the introduction in the system of extra two-level systems which behave as sensors, as described by the Hamiltonian 
\begin{equation}
\label{eq:Hsensor}
{{H}_S}=\sum_s \omega_s\xi _s^{\dagger }{{\xi }_s^-}+\varepsilon \sum_s \left( E \xi_s^-+E^\dagger \xi_s^\dagger  \right),
\end{equation}
with $\xi_s$ ($\xi_s^{\dagger}$) the lowering (rising) operator for sensor $s$, and $\omega_s$ its resonant frequency, in the rotating frame at the laser frequency~\footnote{We here consider sensors which all couple to the field radiated in the same direction, but a generalization to two-direction photon-photon correlations can be obtained by introducing sensors which couple to the field \eqref{eq:E} emitted in different directions.}. The $\varepsilon $ parameter corresponds to the coupling strength between the sensors and the atomic system, which must be made sufficiently small to not perturb significantly the dynamics of the latter and to avoid the saturation of the sensor ($\varepsilon ={{10}^{-4}}$ throughout the paper).
The sensors are also characterized by their linewidth $\Gamma_s$, a parameter of importance as we shall see later, which manifests in an extra Lindblad term:
\begin{equation}
    \mathcal{L}_S\rho =\frac{\Gamma_s}{2}\sum_s\left(2\xi_s^-\rho \xi_s^\dagger -\xi_s^\dagger\xi_s^-\rho -\rho \xi_s^\dagger\xi_s^-\right).\label{eq:Lsensor}
\end{equation}
The sensor contributions (\ref{eq:Hsensor}-\ref{eq:Lsensor}) are then added to the master equation \eqref{mast}, with $\rho$ now describing the density matrix of the whole system (atoms plus sensors).

The steady-state two-photon time- and frequency-resolved correlation is then obtained from the second-order correlation function from the sensors operators:
\begin{align}
\label{sensortau}
&g_s^{(2)}( \omega_1, \omega_2,\tau  )=\\ \nonumber
&\lim_{t\to\infty}\frac{\left\langle \xi _{1}^{\dagger }( \omega_1,t  )\xi _{2}^{\dagger }(\omega_2,t+\tau  ){{\xi }_{2}}( \omega_2,t+\tau  ){{\xi }_{1}}( \omega_1, t  ) \right\rangle }{\left\langle \xi _{1}^{\dagger }( \omega_1, t ){{\xi }_{1}}( \omega_1, t  ) \right\rangle \left\langle \xi _{2}^{\dagger }(\omega_2,t  ){{\xi }_{2}}(\omega_2,t  ) \right\rangle }.
\end{align}
Equal-time correlations ($\tau=0$) characterize the simultaneous emission of photons of frequencies $\omega_1$ and $\omega_2$, they are hereafter called 2PFC (two photon frequency resolved correlations) and noted by $g_s^{(2)}( \omega_1, \omega_2)$, for simplicity. Thus, at the expense of two extra degrees of freedom, equal-time frequency-resolved correlations $g_s^{(2)}( \omega_1, \omega_2)$ are contained in the steady-state values of the density matrix, while time- and frequency-resolved ones (i.e., $g_s^{(2)}( \omega_1, \omega_2, \tau)$) are obtained as two-time correlators, using the ``standard'' (two-time) quantum regression theorem~\cite{Gardiner_2014}. 

 Experimentally, frequency-resolved signals can be obtained using frequency filters such as a Fabry-Perot cavities whose resonance frequency and linewidth correspond to the sensor ones, $\omega_s$ and $\Gamma_s$, but also from time-resolved measurements using beatnote techniques for the $g^{(1)}(\tau)$ function~\cite{OrtizGutirrez2019,Ferreira2020}, for example. Throughout this work, the different correlation functions were computed using the Qutip toolbox~\cite{Johansson2012,Johansson2013} and the Matlab$^{\circledR}$ software, using a solver to reach the steady-state.

\section{Strongly interacting atoms}

\subsection{Fluorescence spectrum}

The radiation spectrum of a strongly driven two-level emitter has a rather intuitive interpretation in the dressed state picture: after the light modes were traced out to obtain Eqs.(\ref{Ham}-\ref{Lin})~\cite{Lehmberg1970a,Lehmberg1970b}, in this picture the photon number is restored to obtain hybrid atom-field states. The resulting atom-field eigenstates have been discussed extensively for single emitters~\cite{Compagno}, and the coupling of light to atom leads to  the following eigenstates for the Hamiltonian at resonance
\begin{equation}
    \left| \pm  \right\rangle =\frac{1}{\sqrt{2}}\left( \left| \uparrow ,n-1 \right\rangle \pm \left| \downarrow ,n \right\rangle  \right),
\end{equation}
where $\left| \downarrow  \right\rangle$  and $\left| \uparrow  \right\rangle$ denote the single-atom ground and excited states, respectively, and $n$ is the photon number in the driving field (i.e.,  the laser). This pair of eigenstates forms the $n$-excitation manifold: in each manifold the eigenstates are split by the Rabi frequency of the driving field (unless the Cavity Quantum Electrodynamics regime is reached~\cite{Jaynes1963,Brune1996}). 

\begin{widetext}

\begin{figure}  %for figs label at side  \sidesubfloat
	\centering 
		\topinset{\bfseries(a)}{\includegraphics[width=0.45\textwidth]{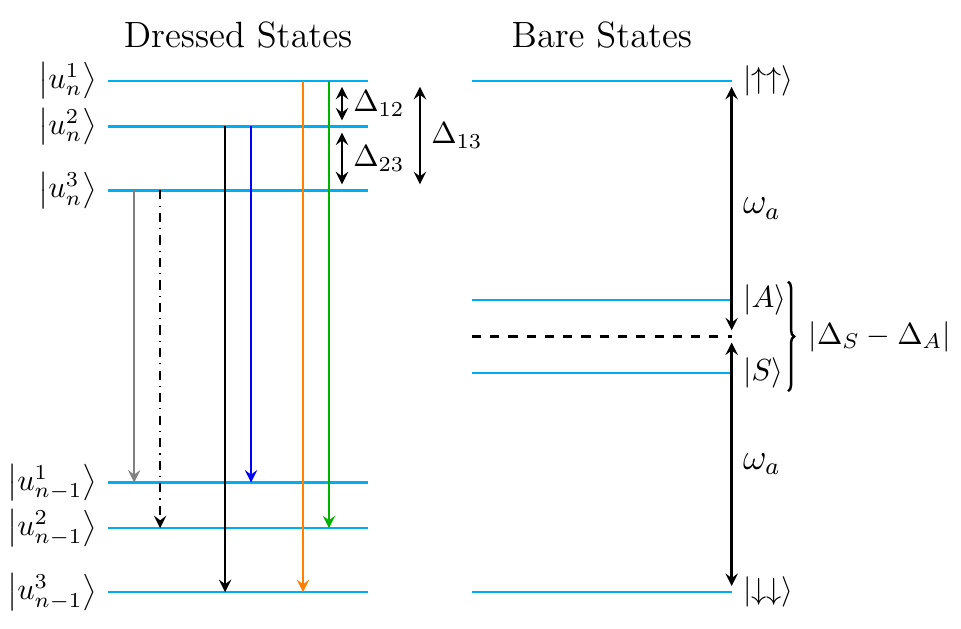}}{-0.1in}{0in}
		\topinset{\bfseries(b)}{\includegraphics[width=0.42\textwidth]{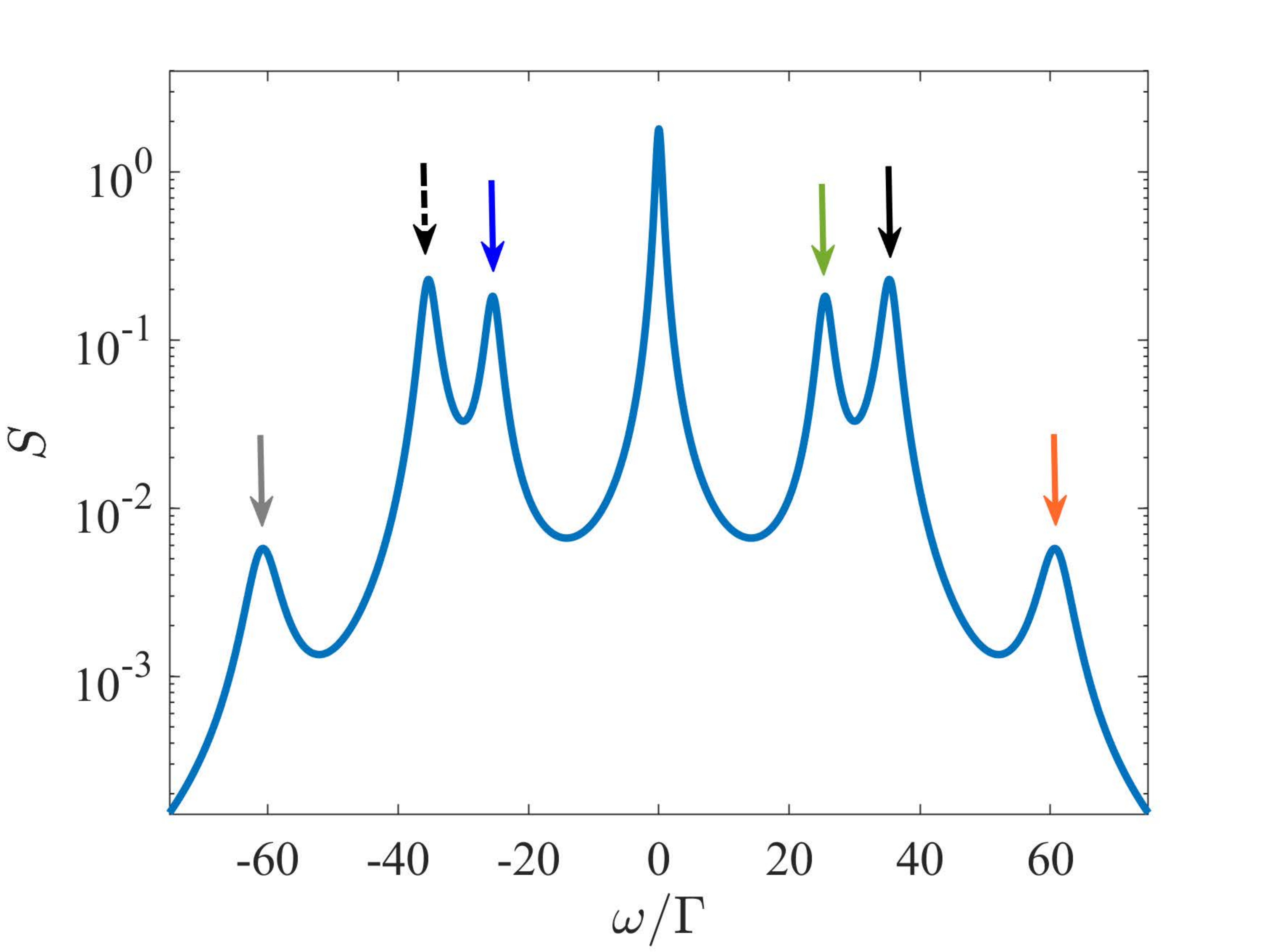}}{0.1in}{0.1in}
	\caption{(a) Collective dressed (left) and bare (right) states for two strongly-interacting atoms, in the rotating frame of the laser and in the lab frame, respectively. (b) 1PS of two interacting atoms. The colored arrows above the peaks correspond to the transitions depicted in (a). Simulations carried out for two atoms driven with a field of Rabi frequency $\Omega =30\Gamma$, separated by a distance $kr_{12}=0.05$, with dipole orientation $\theta_{12} =\cos^{-1}(1/\sqrt{3})$.}
	\label{fig:1}
\end{figure}

\end{widetext}

For a pair of atoms, the dipole-dipole interaction in Eqs. (\ref{Ham}-\ref{Lin}) generates two collective single-excitation eigenstates, labelled symmetric and anti-symmetric:
\begin{eqnarray}
\ket{S}&=&(\ket{\uparrow\downarrow}+\ket{\downarrow\uparrow})/\sqrt{2},\nonumber
\\ \ket{A}&=&(\ket{\uparrow\downarrow}-\ket{\downarrow\uparrow})/\sqrt{2},
\end{eqnarray} 
which present linewidths $\Gamma_S=\Gamma(1+\gamma_{12})$ and $\Gamma_A=\Gamma(1-\gamma_{12})$, and energy shifts $\Delta_S=\Gamma\delta_{12}$ and $\Delta_A=-\Gamma\delta_{12}$. Throughout this work we have fixed $\cos(\theta_{12}) = 1/\sqrt{3}$, which implies $\delta_{12}<0$ for the interatomic distances considered and, consequently, $\Delta_S <\Delta_{A}$, as shown in Fig. \ref{fig:1}.

We here consider the case of two very close, strongly interacting atoms ($kr_{12}\ll 1$ or, more specifically, $\left| {{\delta }_{12}} \right| \gg 1 $ and ${{\gamma }_{12}}\approx 1$), in the presence of a strong resonant driving, characterized by $\Omega^2 > \Gamma^2+4 |\Gamma\delta_{12}|^2$. 
Following the approach of Ref. \cite{Compagno}, we consider the following basis
\begin{subequations}
    \begin{align}
        \left| \phi _{n}^{1} \right\rangle &=\left| \uparrow \uparrow ,n-2 \right\rangle,\\
        \left| \phi _{n}^{2} \right\rangle &=\left| S,n-1 \right\rangle,\\
        \left| \phi _{n}^{3} \right\rangle &=\left| A,n-1 \right\rangle,\\
        \left| \phi _{n}^{4} \right\rangle &=\left| \downarrow \downarrow ,n \right\rangle.
    \end{align}
\end{subequations}
which is the four dimensional subspace of the eigenvectors of the operator ${{N}_{T}}=N_\nu+ 1/2 + \sum\limits_{i=1,2}(1+\sigma _{i}^+\sigma _{i}^-)/2$, where $N_\nu$ is the photon number operator for eigenvalue $n$.
In this basis, the eigenstates of the atom-light system are composed by the collective dressed states incorporating the eigenstates of  Hamiltonian \eqref{Ham} for two atoms with light-mediated interactions, and the photon number states of the light field i.e., the $n$-excitation manifold for our system is given by:
\begin{subequations}
    \begin{align}
\label{moderate}
& \left| u_{n}^{1} \right\rangle =a_1\left| \uparrow \uparrow ,n-2 \right\rangle +a_2\sqrt{2}\left| S,n-1 \right\rangle +a_1\left| \downarrow \downarrow ,n \right\rangle, \\ 
& \left| u_{n}^{a} \right\rangle =\left| A,n-1 \right\rangle, \\
& \left| u_{n}^{2} \right\rangle =-\frac{1}{\sqrt{2}}\left| \uparrow \uparrow ,n-2 \right\rangle +\frac{1}{\sqrt{2}}\left| \downarrow \downarrow ,n \right\rangle,  \\ 
& \left| u_{n}^{3} \right\rangle =a_2\left| \uparrow \uparrow ,n-2 \right\rangle -a_1\sqrt{2}\left| S,n-1 \right\rangle +a_2\left| \downarrow \downarrow ,n \right\rangle,  
    \end{align}
\end{subequations}
 with $a_{1}$ and $a_{2}$ two coefficients obtained by diagonalization of the Hamiltonian (the lengthy expressions for $a_1$ and $a_2$ are not shown here, and the normalization of the eigenstates imposes $a_{1}^{2}+a_{2}^{2}=1/2$).

These dressed states are characterized by an entanglement between atomic and field states, apart from the one containing the antisymmetric atomic state, $ \left| u_{n}^{a} \right\rangle$. Since the latter state is not entangled with the field states, nor is it coupled to the other states through the Hamiltonian, it does not participate to the dressing. 
Furthermore, in the limit of strong coupling of the atoms considered here, it can be shown that this anti-symmetric state does not participate substantially to the steady-state fluorescence spectrum. 
Although it is not driven directly by the laser ($kr_{12}\ll 1$  leads to a rather homogeneous phase profile of the laser on the atoms, thus addressing the symmetric state), it gets substantially populated by decay from the atomic state $\ket{\uparrow\uparrow}$ and its long lifetime allows it to hold a substantial population~\cite{Cipris2020}.  Nevertheless, the weak linewidth also translates in a low number of emitted photons. Thus, unless specified (see Sec.\ref{sec:PPC}), we hereafter neglect this state in our analysis.

 This leads us to introduce the collective operator $\sigma _S^{+- }=(\sigma _{1}^{+- }+\sigma _{2}^{+-})/\sqrt{2}$ and simplify the Linbladian \eqref{Lin} into ${\mathcal{L}_{\sigma_S}}\rho ={\Gamma_S }(2\sigma_S^{-}\rho \sigma_S^{+}-\sigma_S^{+}\sigma_S^{-}\rho -\rho \sigma_S^{+}\sigma_S^{-})$ in the strong interaction regime. As a consequence the $n$-excitation manifold reduces to the triplet of $(\ket{u_{n}^{1}}, \ket{u_{n}^{2}}, \ket{u_{n}^{3}}$), with the frequency difference hereafter called:
\begin{equation}
    \Delta_{ij}=-\Delta_{ji}\equiv E_{n}^{i}-E_{n}^{j}.
\end{equation}
The dressed energy levels are then composed of the $n$-excitation manifolds, each composed of the above triplet, and with successive manifolds separated by energy $\omega_L$: the dressed states and the equivalent bare collective energy levels for two interacting atoms are presented in Fig. \ref{fig:1}(a).

The 1PS is obtained by solving the master equation from Eqs. (\ref{mast}-\ref{Lin}) combined with the quantum regression theorem or, equivalently, monitoring the population of a sensor whose resonant frequency $\omega_s$ is tuned. The fluorescence spectrum for two strongly interacting atoms is depicted in Fig.\ref{fig:1}(b), where the different peaks can be interpreted in the dressed state picture. Similarly to the single-atom case, the central peak originates in the $\ket{u_{n}^{i}}\to\ket{u_{n-1}^{i}}$ transitions ($i=1,\ 2,\ 3$), which do not alter the atomic state and are characterized by the emission of a photon at the laser frequency ${{\omega }_{L}}$. 

The transformation of the doublet of states into a triplet of states for the $n$-excitation manifold, due to the interactions, leads to a seven-peak 1PS. The six sidebands are collective, corresponding to resonant frequencies $\pm\Delta_{ij}$ not present for single atoms  (all transitions are hereafter given in the laser frame), and the corresponding transitions $\ket{u_{n}^{i}}\to\ket{u_{n-1}^{j\neq i}}$ are presented schematically in the dressed state picture of Fig.\ref{fig:1}(a).

\subsection{Photon-photon correlations\label{sec:PPC}}

\begin{figure*}[t]
    \centering
		\topinset{\bfseries(a)}{\includegraphics[height=4.4cm]{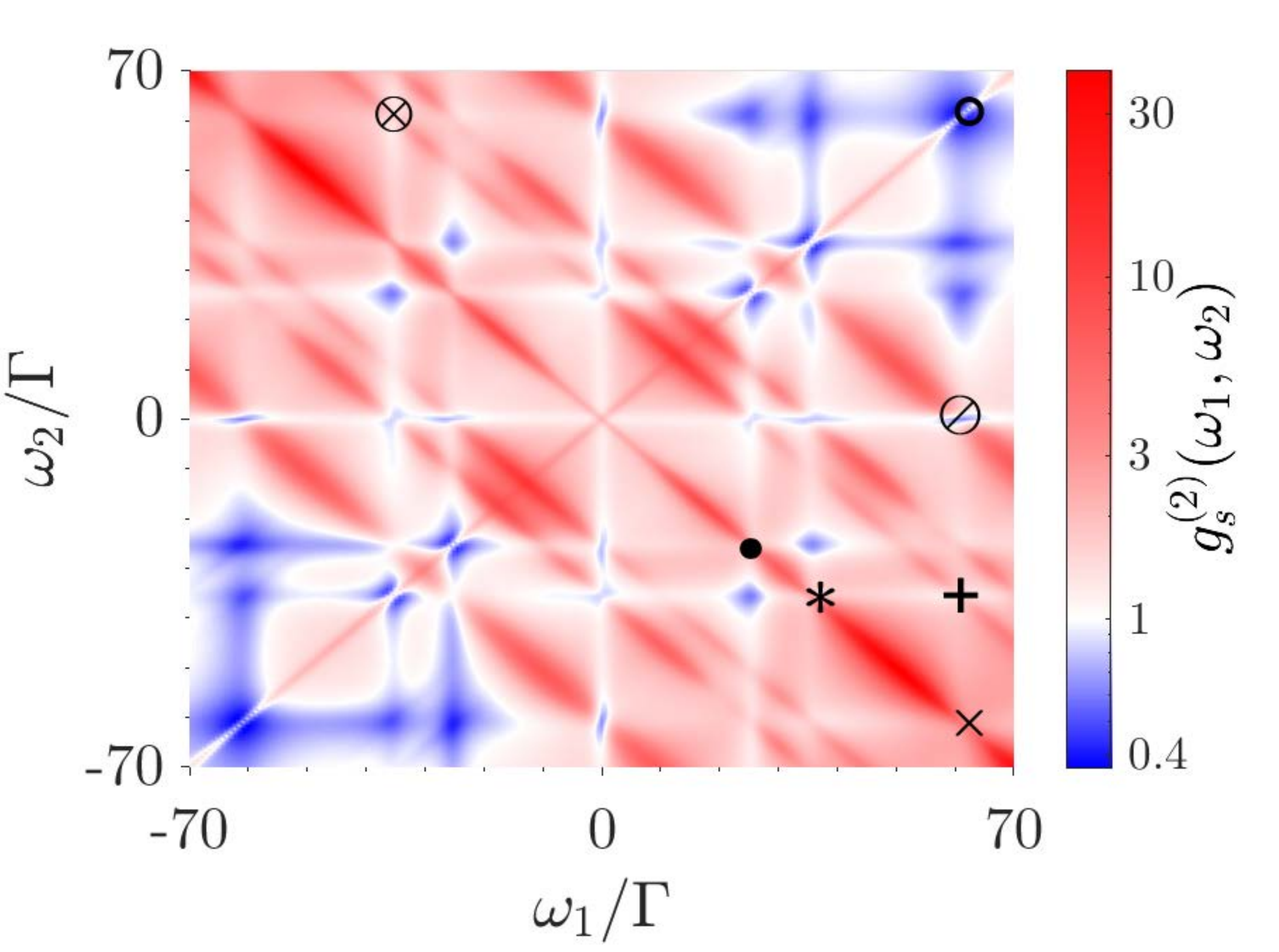}}{0in}{0in}
	    %\hspace{0.1cm}
	    \topinset{\bfseries(b)}{\includegraphics[height=4.4cm]{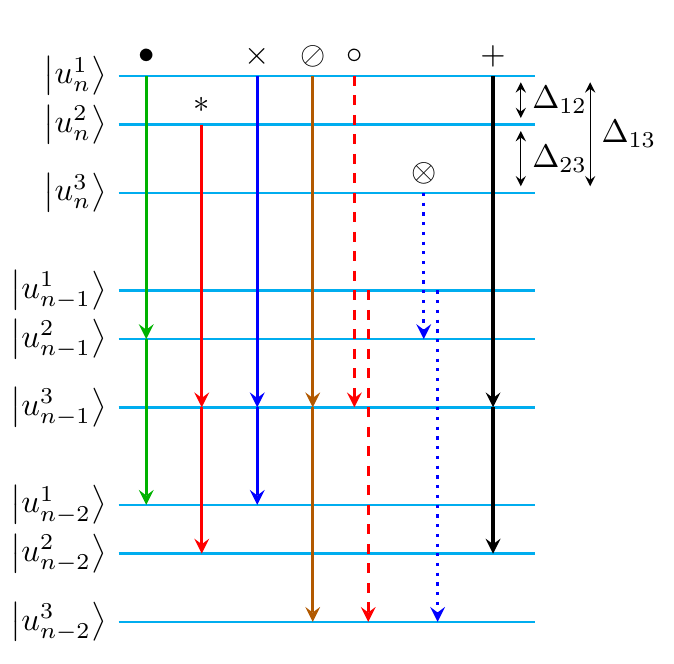}}{0in}{-0.1in}%anewRomain
	    %\hspace{0.1cm}
		\topinset{\bfseries(c)}{\includegraphics[height=4.4cm]{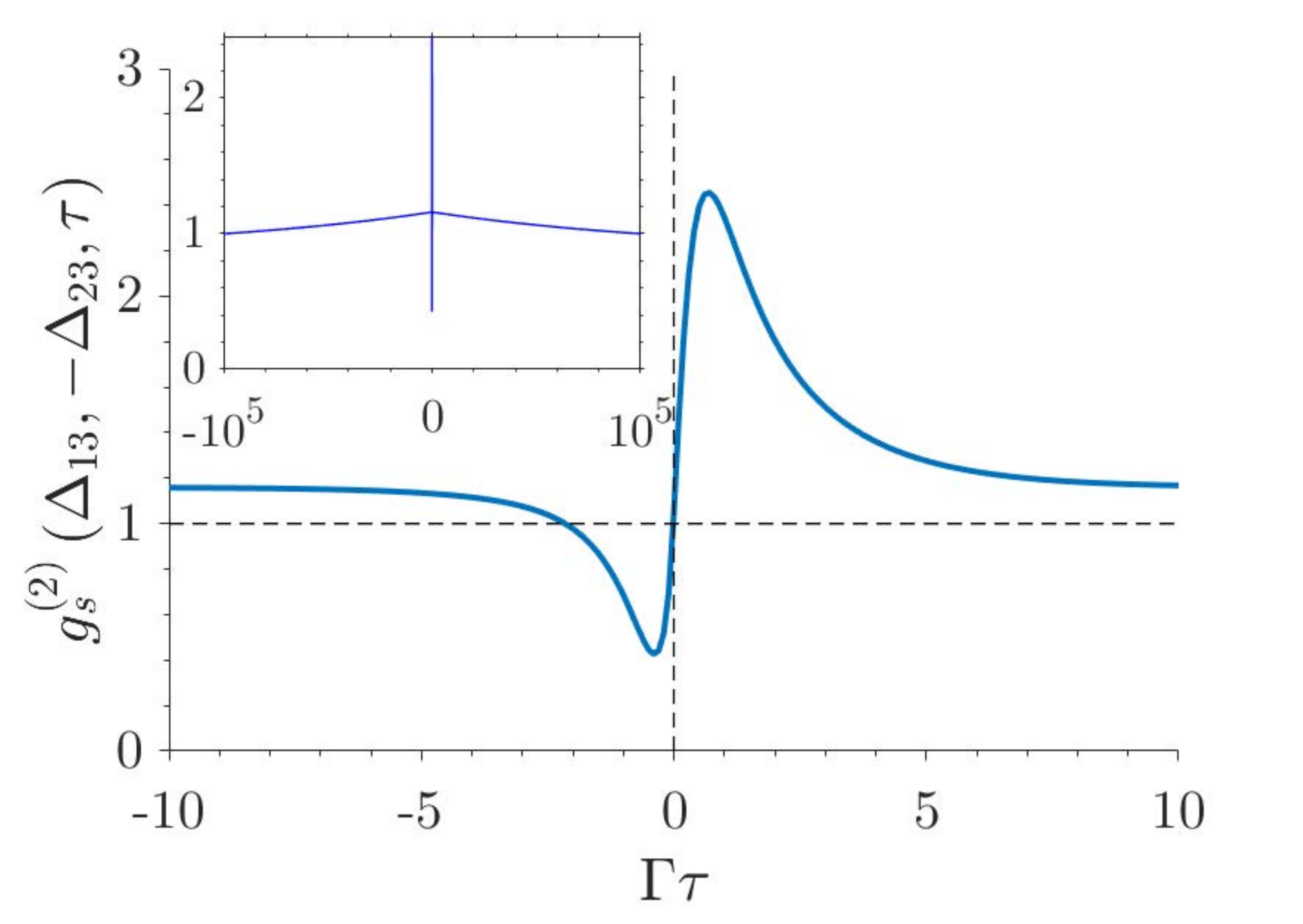}}{0in}{0in}
	\caption{(a) Steady-state photon-photon correlations $g^{(2)}_s(\omega_1,\omega_2)$ for a pair of strongly-interacting strongly-driven atoms. (b) Cascade processes involving the emission of two photons, according to the energy levels of the dressed states picture (allowed cascades with plain lines, forbidden processes with dashed/dotted ones, see text). The associated $g^{(2)}_s(\omega_1,\omega_2)$ is given by the same symbols in (a). (c) Time-resolved $g_s^{(2)}(\omega_1,\omega_2,\tau)$ for the transitions involving photons of frequency $(\omega_1, \omega_2)=(\Delta_{13},-\Delta_{23})$ ( $+$ symbol in (a) and (b)) with $\Gamma_s=\Gamma$. Inset: same curves, for a broader time window. Simulations realized for two atoms with $kr_{12}=0.05$, $\theta_{12}=\cos^{-1}(1/\sqrt{3})$, $\Omega=30 \Gamma$, and $\Gamma_s=\Gamma$ for (a). $kr=0.006$, $\theta=\cos^{-1}(1/\sqrt{3})$, $\Omega=250\Gamma$ and $\Gamma_s=5\Gamma$ for (c).
	\label{fig:2}}
\end{figure*}

Let us now study the specific correlations which occur between these different emission processes. While the transitions from one manifold to the next one are the origin of the 1PS, the correlations in the emitted photons are the essence of the 2PFC, $g_s^{(2)}\left(\omega_{1},\omega_{2}\right)$, computed using Eq.\eqref{sensortau}. The 2PFC corresponding to the situation of Fig.\ref{fig:1} is presented in Fig.~\ref{fig:2}(a), the complexity of which reflects the diversity in photon-photon correlations.

\emph{Opposite-sideband correlations --} We first discuss the correlation between opposite sidebands, i.e., with frequency $+{{\Delta }_{ij}}$ and $-{{\Delta }_{ij}}$, as shown by the $\bullet$ symbol for $(i,j)=(1,2)$ in Fig.~\ref{fig:2}(a): it corresponds to the two-photon cascade $\left| u_{n}^{1} \right\rangle \to \left| u_{n-1}^{2} \right\rangle \to \left| u_{n-2}^{1} \right\rangle$, shown  in Fig.~\ref{fig:2}(b). Being an allowed path of relaxation, it leads to photon-photon bunching, $g_s^{\left( 2 \right)}\left( {{\Delta }_{12}},-{{\Delta }_{12}} \right)>1$: this case is similar to the opposite-sideband bunching effect reported for single emitters~\cite{Schrama,Ulhaq}. 
The same holds true for other transitions of the form $\left| u_{n}^{i} \right\rangle \to |u_{n-1}^{j} \rangle \to | u_{n-2}^{i} \rangle$, 
corresponding to the other sidebands.

\emph{Equal-sideband correlations --} Differently, photons emitted from the same sidebands come antibunched, as in all cases the associated relaxation path is blocked (as long as there is no degeneracy, i.e., $\Delta_{12}\neq\Delta_{23}$). An analogous effect is observed for single atoms. For instance, a photon of frequency ${{\Delta }_{13}}$ automatically leads the system to state $\left| u_{n-1}^{3} \right\rangle$, so the next photon cannot be emitted at the same frequency as it requires for the system to be in a state $\left| u_{n-1}^{1} \right\rangle$ (states $\left| u_{n-1}^{1} \right\rangle$ and $\left| u_{n-1}^{3} \right\rangle$ are orthogonal). For this reason, the associated path of relaxation is considered blocked (see $\circ$ symbol cascade in Fig. \ref{fig:2}(b)), and it is characterized by antibunched photons: $g_s^{\left( 2 \right)}\left( {{\Delta }_{13}},{{\Delta }_{13}}\right)<1$ ($\circ$ symbol in Fig.~\ref {fig:2}(a)). 

Nevertheless, as it can be seen in Fig.~\ref {fig:2}(a),  photons from the same sidebands suffer from being in the ``indistinguishability bunching line'' of the 2PFC. Indeed, two photons with the same frequency cannot be distinguished by the sensor, which in turn leads to bunching effects. This manifests in the ``overbunched''  diagonal line in Fig.\ref{fig:2}(a).

\emph{Cross-sideband correlations --} Let us now discuss processes which involve photons from two different sidebands, corresponding to $g_s^{\left( 2 \right)}\left( \pm {{\Delta }_{ij}},\pm {{\Delta }_{i'j'}} \right)$ with $(i,j)\neq (i',j')$. For these processes which involve the three atomic states, a more careful analysis is needed, as photons of different frequencies may be emitted in a specific order. For instance, the double transition $\left| u_{n}^{1} \right\rangle \to \left| u_{n-1}^{3} \right\rangle \to \left| u_{n-2}^{2} \right\rangle$, indicated by a + symbol in the dressed state representation of Fig.~\ref{fig:2}(b), is allowed, and thus permits the successive emission of photons of frequency $\Delta_{13}$ and $-\Delta_{23}$, in that order. Differently, a photon of frequency $-\Delta_{23}$ cannot be followed by one of frequency $\Delta_{13}$, since this would correspond to the successive $\left| u_{n}^{3} \right\rangle \to \left| u_{n-1}^{2} \right\rangle$ and $\left| u_{n-1}^{1} \right\rangle \to \left| u_{n-2}^{3} \right\rangle$ (see $\otimes$ symbol in Fig.~\ref{fig:2}(b)), which is a blocked path since $\left| u_{n-1}^{2} \right\rangle$ and $\left| u_{n-1}^{1} \right\rangle$ are orthogonal. 

Monitoring the zero-delay photon-photon correlations $g_s^{(2)}(\Delta_{ij},\Delta_{i'j'})$ does not allow to distinguish the two processes, but its time-resolved version does. As illustrated by the computation of $g_s^{(2)}(\Delta_{13}, -\Delta_{23}, \tau)$ in Fig.~\ref{fig:2}(c), we observe a strong bunching at delays $\tau\sim+1/\Gamma$, but a below-unity $g_s^{(2)}$ for $\tau\sim -1/\Gamma$ (negative times corresponds to the reverse order, since $g_s^{(2)}(\Delta_{13}, -\Delta_{23}, \tau)=g_s^{(2)}(-\Delta_{23},\Delta_{13},  -\tau))$. The same phenomenon is observed for transitions with photon pairs of frequency $(\pm\Delta_{12},\mp\Delta_{13})$ and $(\pm\Delta_{12},\pm\Delta_{23})$. Thus, these double transitions, which involve the three different atomic states, present a time-symmetry breaking for the $g_s^{(2)}$ function, which corresponds to a specific ordering of the emitted photons. It is due to the interaction between the emitters, which leads to a splitting of the energy levels of the atomic system.

It is interesting to note that on timescales $\tau$ of several single-atom excited state lifetime, the correlator $g_s^{(2)}(\Delta_{13},-\Delta_{23},\tau)$ does not go to $1$ as one would expect: this is the signature of the anti-symmetric state holding a substantial part of the atomic excitations, yet these are released on the mode timescale~\cite{Cipris2020} (see inset of Fig.\ref{fig:2}(c)).

Finally, the case of double processes that depart twice from the same atomic state, but goes to the two others atomic states (i.e., $\left| u_{n}^{i} \right\rangle \to |u_{n-1}^{j}\rangle$ and $\left| u_{n-1}^{i} \right\rangle \to | u_{n-2}^{l} \rangle $, with $i$, $j$ and $l$ all different), are naturally anti-bunched. Indeed both possible orders for the double transition are blocked. Consequently, $g_s^{2}(\Delta_{ij}, \Delta_{il})$ is below unity, as can be observed in Fig.\ref{fig:2}(a).

\emph{Sideband-central peak correlations --} Finally, cascades which involve one sideband photon plus a central peak photon can in principle occur successively, since the latter does not involve a change in the atomic state (i.e., $\left| u_{n}^{i} \right\rangle \to \left| u_{n-1}^{i} \right\rangle $). Furthermore, both orders of emission for the photons could equally occur.
Nonetheless, these pairs of photons come anti-bunched. As discussed for the case of single emitters~\cite{Arnoldus_1984}, this effect originates from a destructive interference, due to the fact that the state of the system is not modified by Rayleigh emission. 
Thus, despite the two cascades involving photons of frequency ${\Delta_{ij}}$ and $0$ are degenerate (they have the same initial and final states), the interference between the amplitude of their transition probability prevents the process instead of favoring it (see Fig.\ref{fig:2}(a)). 

Finally, we point out that a clear observation of antibunching, and other kinds of photon-photon correlations as we investigate in this paper, requires the use of sensors of linewidth at least comparable to the atomic linewidth (they are here taken equal: $\Gamma_s=\Gamma$). Indeed, it has recently been shown that antibunching (and other photon-photon correlations), whether in the temporal domain~\cite{LopezCarreo2018,Hanschke2020,Phillips2020} (as given by $g^{(2)}(\tau)$) is strongly reduced in case of a sublinewidth filtering, since it results in long integration time, which in turn averages out the correlations~\cite{munoz2014violation}.

\begin{figure*}[t]
\centering
		\topinset{\bfseries(a)}{\includegraphics[height=4.6cm]{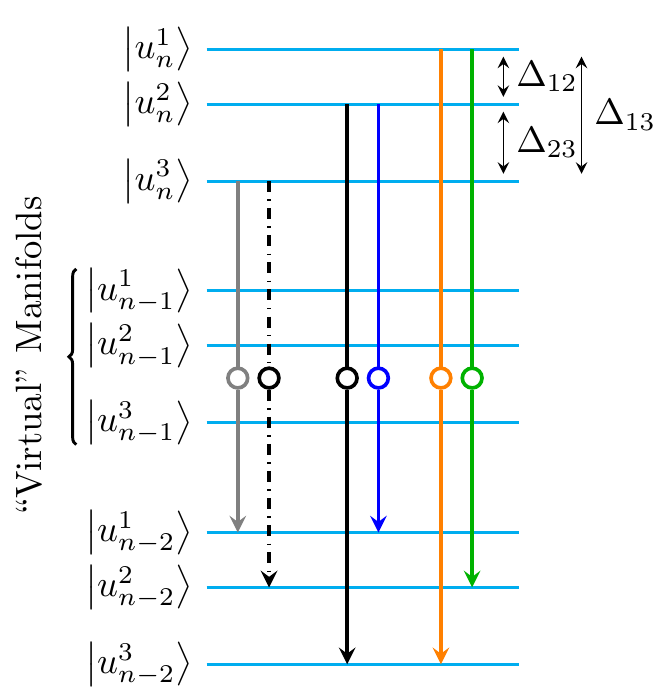}}{0in}{0in}%2aaRomain.pdf
		\topinset{\bfseries(b)}{\includegraphics[height=4.6cm]{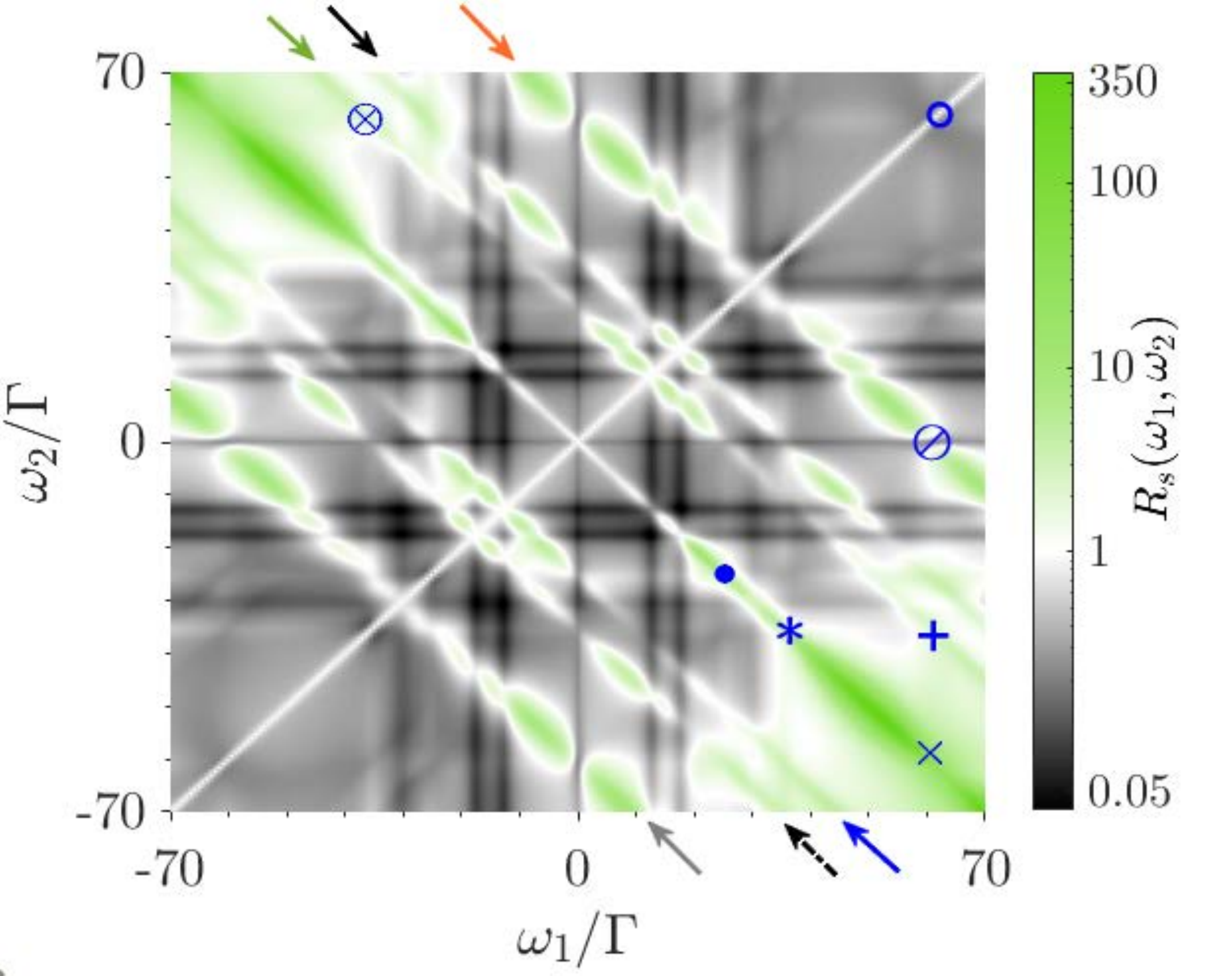}}{0in}{0in}
		\topinset{\bfseries(c)}{\includegraphics[height=4.6cm]{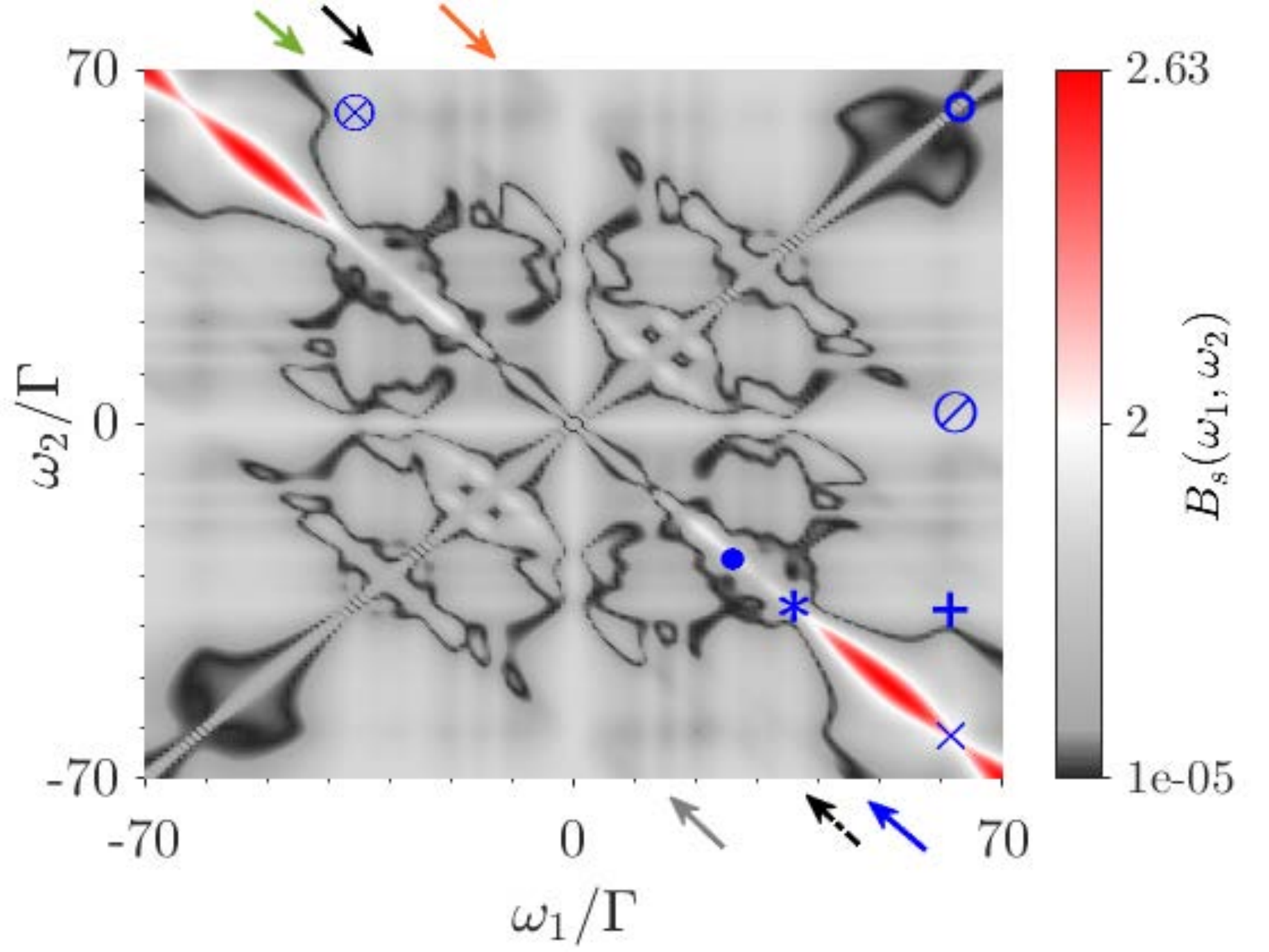}}{0in}{-0.05in}
\caption{(a) Two-photon ``leapfrog'' processes, with the system transiting through virtual dressed levels. (b) Ratio $R_s$ from the Cauchy-Schwartz inequality, and (c) $B_s$ from the Bell inequality, when tuning the frequency of each sensor. Simulations realized for $kr=0.05$, $\theta=\cos^{-1}(1/\sqrt{3})$, $\Omega=30\Gamma$ and $\Gamma_s=\Gamma$.
\label{fig:3}}
\end{figure*}

\subsection{Leapfrog processes}

The cascades described above involve two-photon emission processes that encompass real transitions through intermediate states, where the system state is described by the manifolds from the dressed atom picture. There exist other kinds of two-photon  transitions, where the system does not transit through one of these intermediate states, but rather through a ``virtual'' manifold, labelled ``leapfrog processes''~\cite{Gonzalez}. These transitions are characterized by the joint emission of two photons, and have recently been observed in single quantum dots~\cite{Peiris}. Most of these two-photon collective processes yield correlations much stronger than real transition ones, and their quantum nature has been demonstrated for single emitters using Cauchy-Schwartz and Bell inequalities~\cite{munoz2014violation}.

For these leapfrog processes, the energy of each photon does not need to be related to a specific level transition energy, only their sum needs to obey the following relation:
\begin{equation}
\label{eq.leap1}
\omega_1+\omega_2=0,\ \pm {{\Delta }_{ij}},
\end{equation}
where the frequency in the laboratory frame is obtained by adding $2\omega_L$.
% Elnaz: since each frequency needs $\omega_L$ for lab frame, $\omega_1+\omega_2$ needs $2\omega_L$  for lab frame...............
The leapfrog transitions correspond to the anti-diagonal lines marked by color arrows in Fig.~\ref{fig:3}(b) and (c), and the associated (virtual) transitions are depicted in Fig.~\ref{fig:3}(a). Note that if, in addition to condition \eqref{eq.leap1}, the energy of each photon belongs to the allowed real transitions, the photon emission process is that described in the previous section, and the correlations between the photons are classical.

To characterize the non-classicality of these correlations, we use the Cauchy-Schwartz inequality (CSI) for the second-order correlation functions at zero delay $g^{(2)}_{kl}=g^{(2)}_s(\omega_k,\omega_l)$:
\begin{equation}
    \left[g^{(2)}_{12}\right]^2\leq g^{(2)}_{11} g^{(2)}_{22},
\end{equation}
which we monitor by studying the ratio 
\begin{equation}
    R_s=\frac{\left[g^{(2)}_{12}\right]^2}{g^{(2)}_{11} g^{(2)}_{22}}.
\end{equation}
Values $R_s$ larger than unity are the signatures of non-classical correlations between the two emitted photons~\cite{munoz2014violation}.

\begin{figure*}[t]
\centering
		\topinset{\bfseries(a)}{\includegraphics[height=4.1cm]{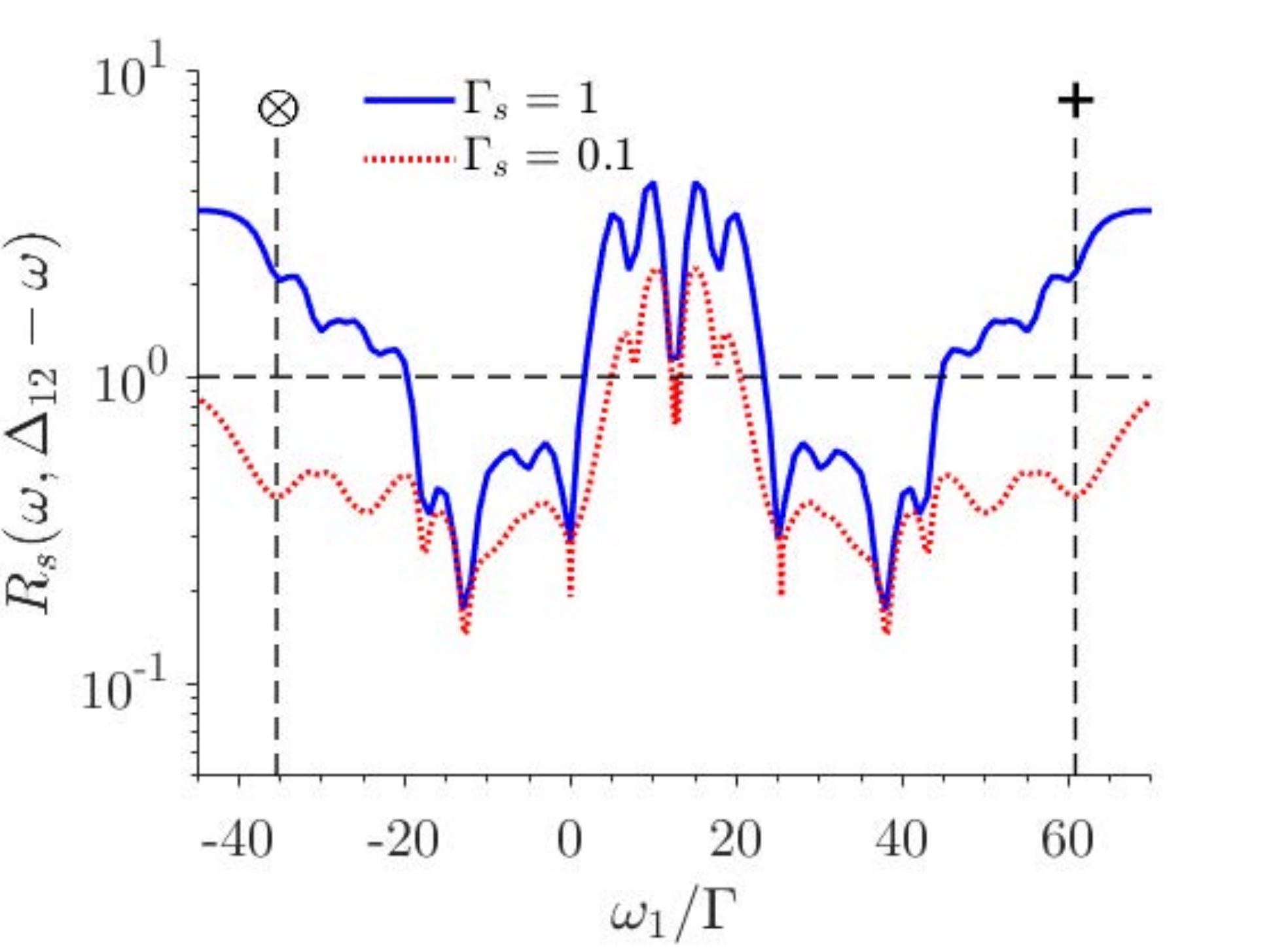}}{0in}{0in}%2aaRomain.pdf33
		\topinset{\bfseries(b)}{\includegraphics[height=4.1cm]{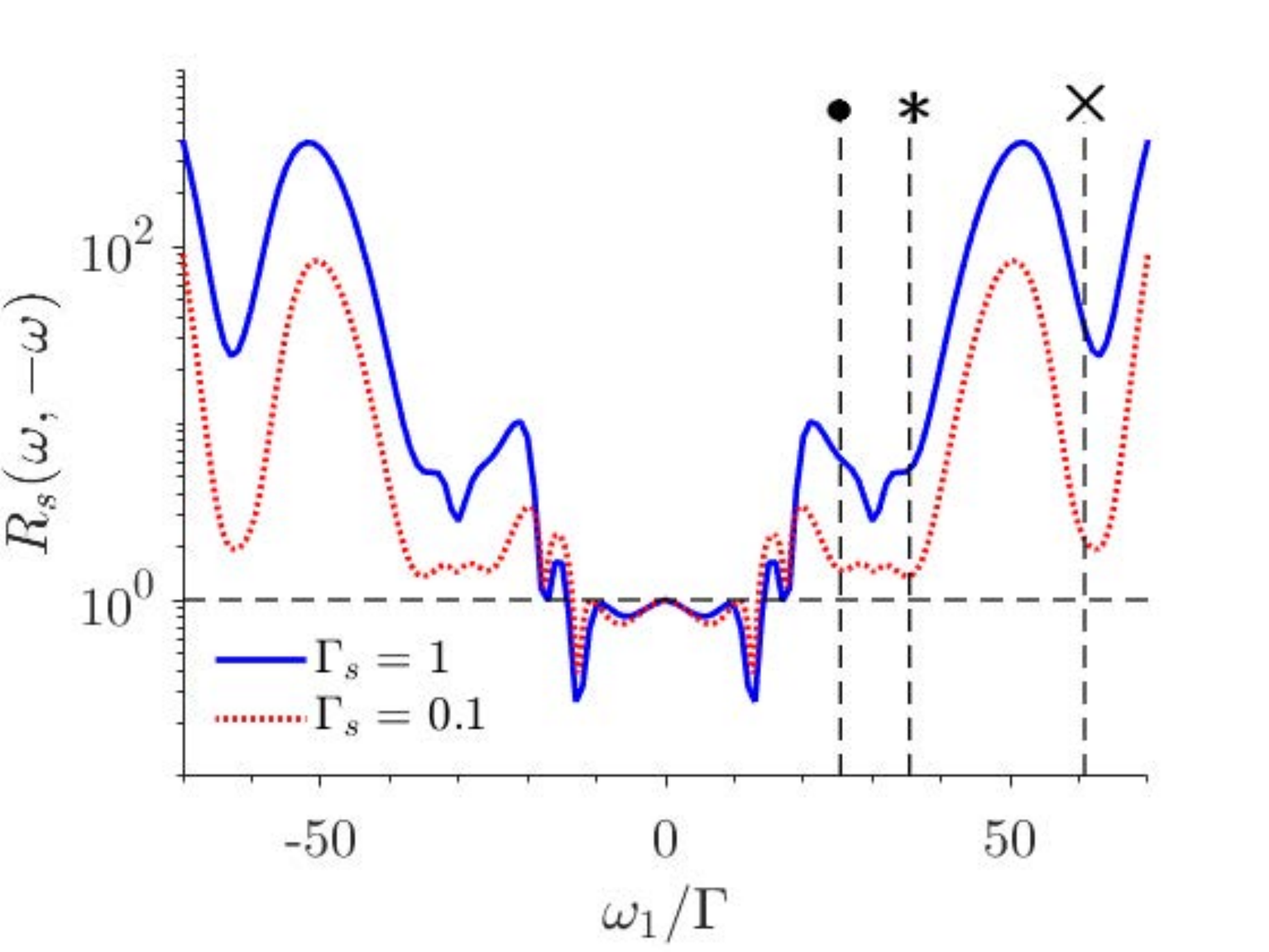}}{0in}{0in}
		\topinset{\bfseries(c)}{\includegraphics[height=4.1cm]{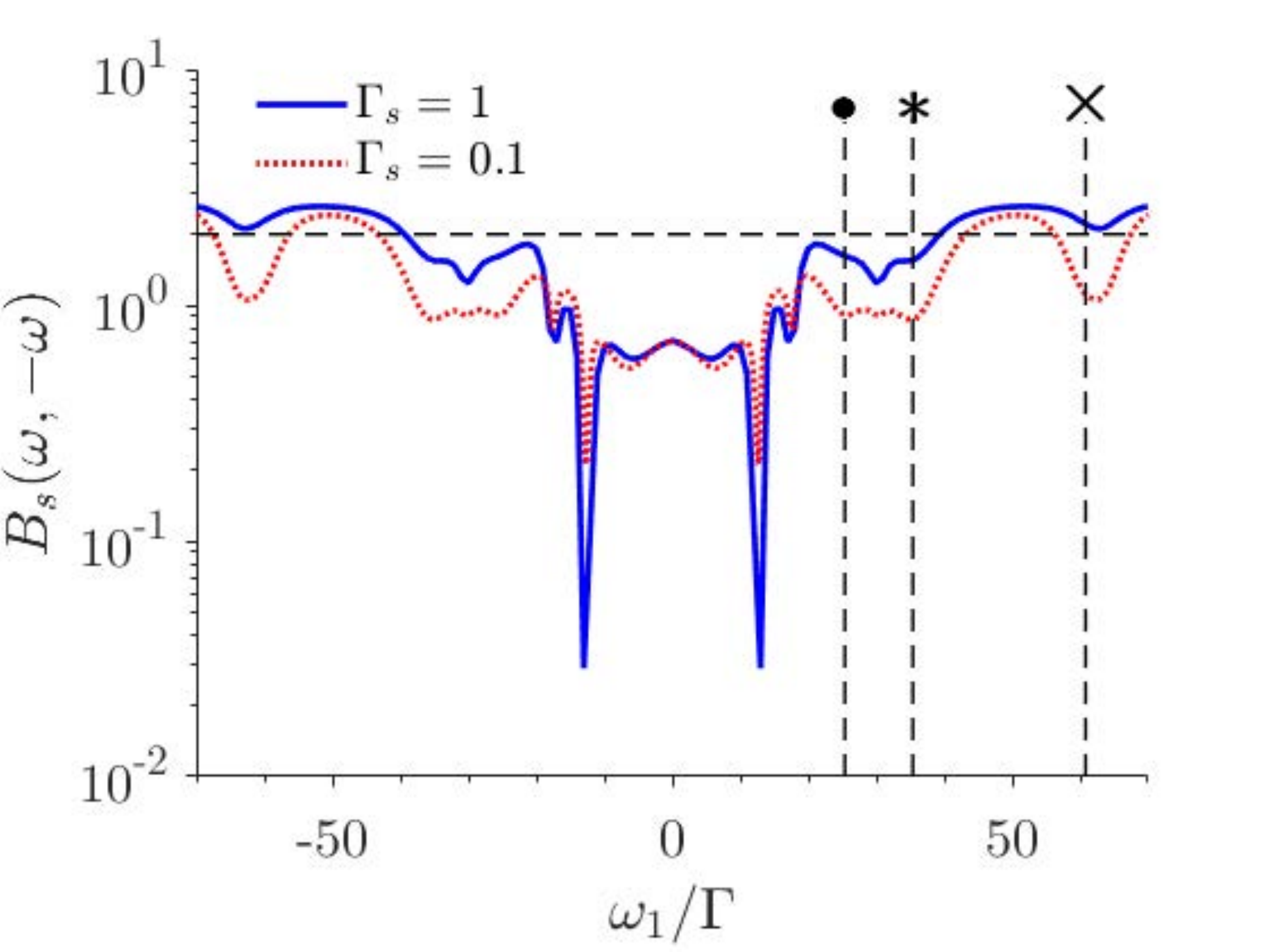}}{0in}{-0.05in}
\caption{Quantifier $R_s$ for the Cauchy-Schwartz inequalities (with a violation above the dashed line) for ${{\Gamma }_{s}}=\Gamma$ (solid blue) and ${{\Gamma }_{s}}=\Gamma/10$ (dotted red), for the leapfrog lines (a) ${{\omega }_{1}}+{{\omega }_{2}}={{\Delta }_{12}}$ and (b) ${{\omega }_{1}}+{{\omega }_{2}}=0$. (c) Quantifier $B_s$ for the Bell inequalities (with a violation above the dashed line) for ${{\Gamma }_{s}}=\Gamma$ (solid blue) and ${{\Gamma }_{s}}=\Gamma/10$ (dotted red) in the leapfrog line of ${{\omega }_{1}}+{{\omega }_{2}}=0$. The vertical lines with a symbol at the top refer to the processes discussed in Figs.\ref{fig:2}(a) and \ref{fig:3}(b) and (c). Simulations realized for $kr=0.05$, $\theta=\cos^{-1}(1/\sqrt{3})$ and $\Omega=30\Gamma$.}
\label{fig:4}
\end{figure*}

In Fig.\ref{fig:3}(a), leapfrog processes which involve different initial and final atomic states are presented: the system does not emit photons from specific (``real'') transitions, only the sum of the two photon energies corresponds to a transition between specific states in the dressed states picture. 
As one can observe from the anti-diagonal lines in Fig.\ref{fig:3}(b), which correspond to $\omega_1+\omega_2=0,\ \pm {{\Delta }_{ij}}$, the CSI is violated for most of these joint emission processes ($R_s>1$). Nevertheless, the CSI is not violated for antibunched real transitions, i.e., for photon pairs with frequencies $(0,\pm{{\Delta }_{ij}})$ or $(\pm{{\Delta }_{ij}},0)$ (see, for example, the $\oslash$ for frequencies $({{\Delta }_{13}},0)$).
Neither it is for pairs of real photons, for frequencies $(\pm{{\Delta }_{12}},\pm{{\Delta }_{23}})$. Also, as we can observe in Fig.~\ref{fig:4}(a-b), a sublinewidth filtering leads to weaker violations of CSI~\cite{munoz2014violation}.

Finally, one observes that the inequality is also violated for some emission processes which involve real transitions, where the correlations between emitted photons are classical (see $+$ or $\otimes$ symbols, for example, in Fig.\ref{fig:3}(b)). 
In order to properly observe the classicality
of these correlations, as they correspond to photons of real transitions located in the leapfrog lines, it is necessary to use sensor with a better resolution. Indeed,
the use of a sensor linewidth $\Gamma_s = \Gamma$ leads to an averaging over processes with different kinds of correlations. To illustrate this point, we show in Fig.~\ref{fig:4}(a) the change in CSI as the sensor linewidth is changed from $\Gamma$ to $\Gamma/10$: the above-mentioned transitions for the real photons do not violate any longer CSI, showing the classical nature of their correlations~\cite{munoz2014violation}.

Furthermore, as for single emitters~\cite{munoz2014violation}, violations of CSI may appear for transitions of the central antidiagonal line ($\omega_1+\omega_2=0$), even for real transitions and for well-resolved frequencies: Fig.~\ref{fig:4}(b) presents such violations of CSI (see $\bullet$,  $\ast$ and $\times$ symbols). This failure of CSI to detect classical correlations can be addressed by using Bell inequality (BI), as monitored by a quantifier adapted to the sensor approach~\cite{munoz2014violation}:
\begin{equation}
B_s = \sqrt{2} \left| \frac{B_{1111} +B_{2222} - 4B_{1221} - B_{1122} - B_{2211}  }{B_{1111} + B_{2222} + 2B_{1221}} \right|,
\end{equation}
with $B_{jklm} = \langle \xi_1^{\dagger}\left(\omega_j\right) \xi_2^{\dagger}\left(\omega_k \right)\xi_2\left(\omega_l\right) \xi_1\left(\omega_m \right) \rangle$.
Values $B_s>2$ are a violation of the BI, which are considered as a true signature of quantum correlations. As can be seen in Fig.~\ref{fig:3}(c) for ${{\Gamma }_{s}}=\Gamma$, BI is violated only for specific areas of the central antidiagonal line, yet not for the real transitions of $\bullet$ and $\ast$. This behaviour is similar to the single-emitter case~\cite{munoz2014violation}, confirming that only transitions involving virtual states hold true quantum correlations between the emitted photons. Note that BI and CSI are sensitive to the frequency resolution of the sensors, as narrow linewidth sensors correspond to long averaging time, which in turn washes out the correlations~\cite{munoz2014violation}.

As illustrated for the pair of photons $(\pm{{\Delta }_{13}},\mp{{\Delta }_{13}})$, a sensor linewidth $\Gamma_s=\Gamma$ presents a violation of BI, yet reducing the sensor linewidth to $\Gamma/10$ removes the violation of the BI, see Fig.~\ref{fig:4}(c) where the pair of $({{\Delta }_{13}},-{{\Delta }_{13}})$ indicated by $\times$. 
This highlights again the narrow linewidth sensors correspond to a finer structure for the quantum quantifiers and the necessity of using sensors with a linewidth comparable to the atomic transition one, in order to detect the stronger correlations.

\section{Conclusion and perspectives}

Strong interactions between two two-level emitters give rise to a series of new sidebands in the fluorescence spectrum, whose shift from the atomic transition depends on both the interaction strength and the driving field. Similarly to the single emitters, the leapfrog processes with frequencies that sums to zero or the frequency of one of the interaction-induced sidebands, are characterized by strong correlations, which can be either classical or quantum. This suggests that strongly coupled emitters are potential sources of heralded photons, with extra control parameters through the interaction, as compared to single emitters.

Another regime of interest is that of a weak dipole-dipole interaction, i.e., when the collective dressed levels are equally spaced in energy ($\Delta_{12}=\Delta_{23}$), which occurs when the distance between the emitters is comparable or larger than the optical wavelength. In this regime, the first correction to the single-atom fluorescence spectrum is the emergence of sidebands shifted from the laser frequency by twice the Rabi frequency ($\omega=\pm 2\Omega$). This phenomenon was predicted to scale with the resonant optical depth for dilute extended clouds, as a signature of the raising two-atom quantum correlations~\cite{Pucci}. We have monitored photon-photon correlations $g^{(2)}(\omega_1, \omega_2)$ for a pair of atoms at a distance of the order of $\lambda$ ($kr=0.4-1$ and $\theta=\cos^{-1}(1/\sqrt{3})$) and strongly driven, yet photon-photon correlations appear to be largely dominated by single-atom physics. This is reminiscent of the anti-bunching phenomenon which vanishes for a large number of independent emitters, unless specific conditions for their interference is achieved~\cite{Grangier_1986}. Furthermore, the leapfrog processes associated with the new sidebands, $\omega_1+\omega_2=\pm 2\Omega$, present no violation of the Cauchy-Schwartz inequality (not shown here). This suggests that despite these sidebands result from correlations between the quantum fluctuations of the two dipoles~\cite{Pucci}, the photons associated with these processes may be only classically correlated.

The variety of sidebands and photon-photon correlations encountered for a pair of atoms calls for a dedicated study for larger systems. Indeed, although the coherent manipulation of atoms at scales below the diffraction limit is experimentally challenging, schemes have been proposed to surpass these limitations, based on the transparency-induced dark states~ \cite{Agarwal2006,Cho2007,Yavuz2007,Gorshkov2008}, which have already allowed for the generation of subwavelength cold atom structures~\cite{Miles2013,Wang2018,Subhankar2019,Tsui2020}. In this context, strongly-interacting cold atom ensembles may be a promising tunable source for entangled pairs of photons, but also for larger bunches of photons~\cite{LopezCarreno}.

\section*{Acknowledgment}

M.\,H., R.\,B. and C.\,J.\,V.-B. acknowledge funding from the French National Research Agency (projects QuaCor ANR19-CE47-0014-01). E.\,D., R.\,B. and C.\,J.\,V.\,-B. benefited from Grants from S\~ao Paulo Research Foundation (FAPESP, Grants Nos. 2018/10813-2, 2018/01447-2, 2018/15554-5, 2019/13143-0, and 2019/11999-5) and from the National Council for Scientific and Technological Development (CNPq, Grant Nos.\,302981/2017-9, 409946/2018-4, and 307077/2018-7). M.\,H. and R.\,B. received support from the project CAPES-COFECUB (Ph879-17/CAPES 88887.130197/2017-01).

%FAPESP grants:2015/25146-3, 2018/25135-0 (license - Matlab).

%%%%%%%%%%%%%%%%%%%%%%%%%%%%%%%%%%%%%%%%
%% BIBLIOGRAPHY
%%%%%%%%%%%%%%%%%%%%%%%%%%%%%%%%%%%%%%%%
\clearpage
\bibliographystyle{apsrev4-2}
\bibliography{Refs}

\end{document}